\documentclass[pra,preprint]{revtex4}
\usepackage[centertags]{amsmath}
\usepackage[koi8-r]{inputenc}
\usepackage{amsfonts}
\usepackage{amssymb}
\usepackage{amsthm} 
\usepackage{newlfont}
\usepackage{epsfig}
\usepackage{amscd}
\usepackage{footnote}
\usepackage{lipsum}
\usepackage{caption}
\usepackage{subcaption}
\usepackage{color}
\usepackage{xcolor}
\usepackage{multirow}
\usepackage{amscd}

\newcommand{\beq}{\begin{equation}}
\newcommand{\eeq}{\end{equation}}
\newcommand{\ba}{\begin{array}}
\newcommand{\ea}{\end{array}}
\newcommand{\bea}{\begin{eqnarray}}
\newcommand{\eea}{\end{eqnarray}}

\begin{document}

\begin{center}
{\large \sc \bf{One-excitation spin dynamics in homogeneous closed chain governed by $XX$-Hamiltonian}}

\vskip 15pt

{\large 
E.B.Fel'dman, E.I.Kuznetsova, A.I.Zenchuk
}

\vskip 8pt

{\it Federal Research Center of Problems of Chemical Physics and Medicinal Chemistry RAS,
Chernogolovka, Moscow reg., 142432, Russia}.

\end{center}

\vskip 8pt


\begin{abstract}
We analytically investigate the one-excitation spin dynamics in a homogeneous  closed spin-1/2 chain via diagonalization  of the one-excitation block of the XX-Hamiltonian, which allows to derive the analytical expressions for probability amplitudes describing state transfers between any two spins of a chain. 
We analytically investigate the $M$-neighbor approximation ($M\ge 1$) of spin dynamics with arbitrary initial state and analyze its accuracy  using special integral characteristics defined in terms of the above probability amplitudes. We find $M$ providing the required accuracy of evolution 
approximation for chains of different lengths.
\end{abstract}



\maketitle

\section{Introduction}
\label{Section:introduction}

The problem of quantum state transfer \cite{Bose,CDEL,KS,GKMT} is one of the central tasks in quantum informatics. Analytical methods solving the problem are mainly  restricted to approximation of the nearest neighbor interactions (NNI) which allows applying the Jordan-Wigner transformation \cite{JW,CG} thus reducing the system of interacting spins to the system of non-interacting fermions. {|color{red} As a  simplified model allowing nevertheless to exhibit a set of principal features of physical processes involved in spin-state dynamics, the nearest-neighbor approximation attracts strong attention of researchers.  Applying this model yields a set of  interesting results} \cite{Abragam}. In particular, we underline  interpretation of magnetic  resonance experimental results   \cite{ALC,BFV}, one-dimensional multiple-quantum spin dynamics \cite{DMF}, quantum information transfer \cite{CRC2007}, studies on entanglement and discord as measures of quantum correlations \cite{KKWC,YLLF},  quantum state transfer \cite{CDEL,KS,Z_2014} and its robustness to small perturbations \cite{ZASO,ZASO2,ZASO3}.   Among other applications, we especially select the perfect state transfer  along a completely inhomogeneous spin chain with XX-interaction \cite{CDEL,KS} based on the mutually rational eigenvalues (i.e., their ratios are rational numbers) of the non-homogeneous nearest-neighbor Hamiltonian.

The nearest-neighbor approximation is well applicable to the exchange interaction when the coupling constants (exchange integrals) decrease exponentially with the distance between the interacting particles \cite{Yo,LL}. 
As for the systems with the dipole-dipole interaction (DDI), NNI is applicable only over relatively short time intervals when $Dt<1$ \cite{CRC}.  ($D$ is the coupling constant  of  DDI between nearest neighbors). This condition is not satisfied, for instance, in state transfer along the closed (or open)  spin chain involving a large number of spins when the remote-spin interactions become significant. This effect was studied in \cite{FZ_2022,FZ_2023}.

Although numerical methods  are free from the necessity of using NNI, their application is restricted by size of the system under consideration \cite{DFGM}. A partial progress was achieved on the basis of Chebyshev polynomial expansion \cite{DRKH,ZCAPCDRV} and on the effect of quantum parallelism \cite{ADLP}. However, the intensive development of numerical methods do not  allow  describing the full-dimensional dynamics of more then about 20 spins.

Another way to describe the dynamics of large number of spins is restriction to a state subspace with a fixed excitation number, for instance, to the one-excitation state subspace. In this case the dimension of the Hilbert space increases linearly with the length of the chain.  Such dynamics can be observed, for instance,  in spin systems placed in the strong external magnetic field, when the dynamics is governed by the Hamiltonian preserving the excitation number ($z$-projection  of the total spin momentum) \cite{Abragam}. One-excitation spin dynamics was simulated  in set of quoted above papers.  

Here we  consider the  analytical tool for describing the multi-qubit spin dynamics in the one-excitation spin subspace  along the closed homogeneous spin chain governed by the $XX$-Hamiltonian and show that this dynamics   can be described analytically not only for the nearest-neighbor approximation but also for any number $M$ of interacting neighbors. 
The proposed tool is based on the analytical diagonalization  of the one-excitation block  (a symmetric circulant matrix) of the Hamiltonian of the system under consideration \cite{DPh,G}. As a result, we can include  the contribution to the quantum state dynamics from  interactions with all remote nodes.

 Although diagonalization of a circulant matrix has been studied in literature (for instance see  \cite{DPh,G}), we discuss the basic steps of such diagonalization in application to the $N$-qubit spin dynamics of an arbitrary  one-excitation state including interactions of $M$ neighbors ($1\le M \le N-1$, $M$-neighbor approximation; $M=1$ means nearest-neighbor approximation and $N-1$ means all-node interaction).

We apply the $M$-node approximation \cite{FZ_2022}  to the problem of  one-qubit state propagation along the $N$-node closed chain and investigate the accuracy of such approximation via the appropriate integral characteristics \cite{FZ_2022,FZ_2023} taking into account probability amplitudes for all possible  state transfers in the one-excitation  state subspace. 

The paper is organized as follows. In Sec.\ref{Section:diag}, we diagonalize the one-excitation block  of the XX-Hamiltonian in the homogeneous closed chain and construct the evolution operator. The $M$-neighbor approximation  of  quantum state dynamics  in the homogeneous closed spin chain {\it(which is main result of  our paper)} is studied  in Sec.\ref{Section:analyt}.  We show that $M\sim 10$ for the circulant chains of 20-17 nodes. Brief discussion of our results is given in concluding Sec.\ref{Section:conclusions}. 

\section{Diagonalization of one-excitation block of XX-Hamiltonian in a homogeneous closed chain. }
\label{Section:diag}
We consider a homogeneous closed chain ($s=1/2$) governed by the dipole-dipole XX-Hamiltonian which reads
\begin{eqnarray}\label{XX}
H = \sum_{i>j} D_{ij} (I_{x,i} I_{x,j} + I_{y,i} I_{y,j}), \;\; D_{ij} =\frac{\gamma^2 \hbar}{2 r_{ij}^3} (3 \cos^2\theta_{ij} -1),
\end{eqnarray}
where $I_{\alpha,j}$ ($\alpha=x,y,z$) are the operators of the $\alpha$-projection of the spin momentum, $r_{ij}$ is the distance between the $i$th and $j$th spins, $\theta_{ij}$ is the angle between the vector $\vec r_{ij}$ and the external magnetic field, $\gamma$ is the gyromagnetic ratio and $\hbar$ is the Plank constant. 
In what follows we suppose that the magnetic field is perpendicular to the chain plane so that $\theta_{ij}=\pi/2$ for all $i$ and $j$. In addition, for the distance between the nodes on a circular chain of the radius $R$ we have
\begin{eqnarray}
r_{ij} = R \sqrt{2-2\cos\frac{2 \pi |j-i|}{N}}= 2R \sin\frac{\pi |j-i|}{N},
\end{eqnarray}
therefore, 
\begin{eqnarray}
D_{ij} = d_{|i-j|},\;\; d_k = d_{N-k}.
\end{eqnarray}
In addition, if $M$-neighbor approximation is considered, we require
\begin{eqnarray}\label{dk}
d_k=0 \;\;{\mbox{for}}\;\; M<k< N-M.
\end{eqnarray}
Correctness of inequality (\ref{dk}) follows from the fact that $M$ can not exceed $N/2$, see  Eq.(\ref{Nf}) below.
 We shall  notice that, in the case of a one-dimensional chain, Hamiltonian (\ref{XX}) with $M=1$ (nearest neighbor interactions) can be constructed using tool of multiple-quantum Nuclear Magnetic resonance (MQ NMR) \cite{DMF}, where $D_{ij}$ are constants of dipole-dipole interactions.

\begin{figure*}[!]
\includegraphics[scale=0.5]{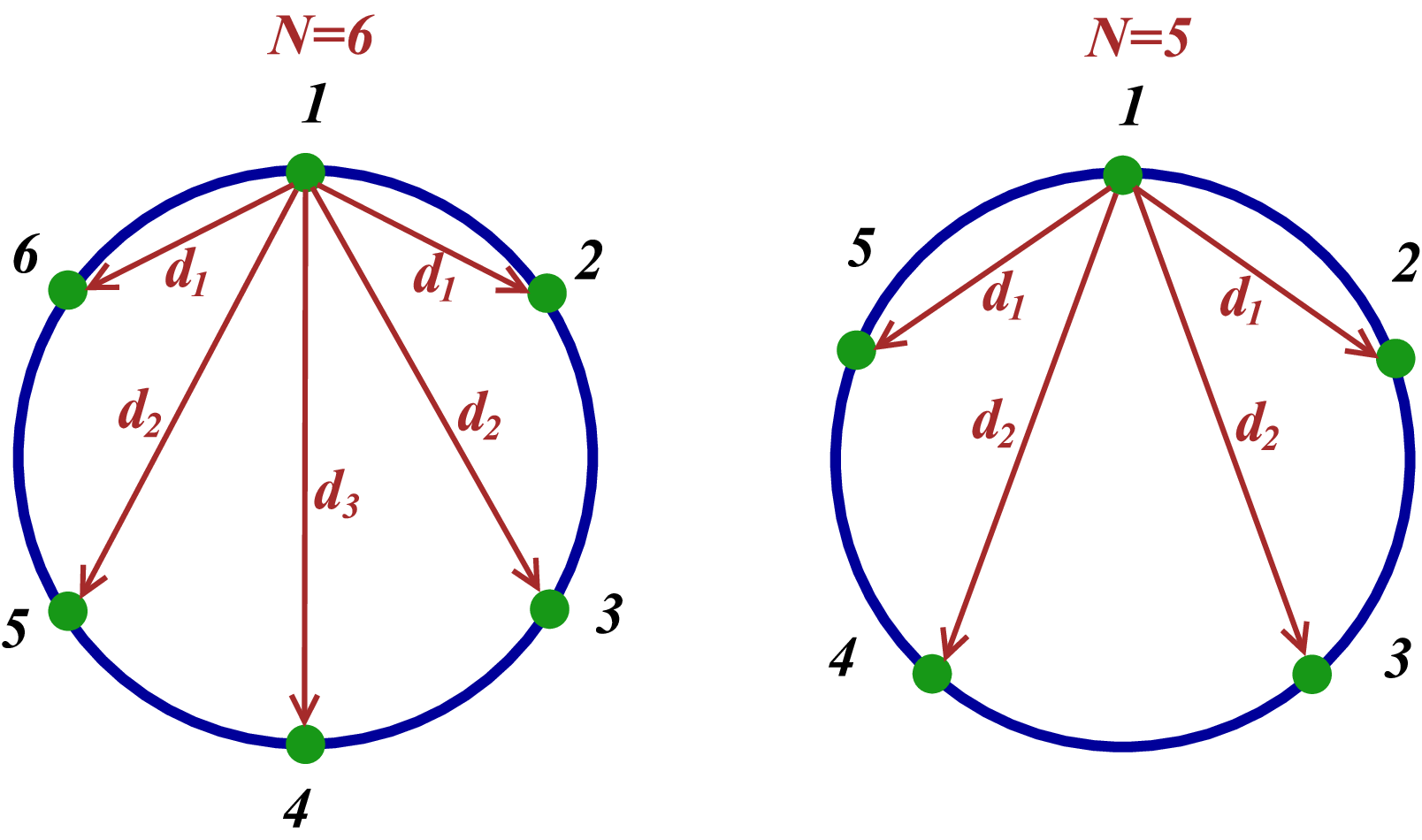}
\caption{Examples of the even- and odd-node closed homogeneous  chains with $N=6$ ($N_f=3$) and $N=5$ ($N_f=2$) respectively. Interactions of the 1st spin with all others  are shown.}
\label{Fig:cycle}
\end{figure*}

Since we study the one-excitation spin dynamics in a homogeneous closed chain our consideration is restricted to the one-excitation block of the Hamiltonian with the following periodic circulant matrix
\begin{eqnarray}\label{HM}
H^{(M)} =\frac{1}{2}\left(
\begin{array}{cccccc}
0  &d_1&d_2&\ddots  &   d_2&     d_1\cr
d_1&0 &d_1&d_2&\ddots   &    d_2\cr
\ddots &\ddots&\ddots&\ddots&\ddots&\ddots \cr
\ddots&d_2&d_1&0  &d_1&d_2\cr
d_2&\ddots&d_2&d_1&0&d_1\cr
d_1&d_2&\ddots&d_2&d_1&0
\end{array}
\right).
\end{eqnarray}
We  discuss the structure of this matrix.  Let  $N_f$ be the  maximal possible number of  interacting neighbors ($M=N_f$ corresponds to all-node interaction),
\begin{eqnarray}\label{Nf}
N_f= \left\{\begin{array}{ll}\displaystyle
{N}/{2},& {\mbox{for even}}\;\;N\cr \displaystyle
({N-1})/{2}&{\mbox{for odd}}\;\;N
\end{array}\right..
\end{eqnarray}
The matrix $H^{(M)}$ at $M<N_f$ has a pair of upper (and a pair of lower) secondary diagonals containing $d_i$, $1\le i \le M$. 
However, we face a difference between the odd and even $N$ if $M=N_f$.
If $N$ is odd then there are two upper secondary diagonals in $H^{(N_f)}$ with $d_{N_f}$ (and two lower secondary diagonals with $d_{N_f}$), while there is only one upper and one lower secondary diagonals with $d_{N_f}$ if $N$ is even, see Fig.\ref{Fig:cycle}.

Let us find the eigenvectors $u_m$ and eigenvalues $\lambda^{(M)}_m$ of $H^{(M)}$ solving the system 
\begin{eqnarray}\label{Hu}
\sum_j H^{(M)}_{kj} u_{mj}=\lambda^{(M)}_m u_{mk},\;\; u_{m(N+1)}= u_{m1}.
\end{eqnarray}
with cyclic condition
\begin{eqnarray}\label{cycle}
\;\;u_{m(k+N)} = u_{mk}.
\end{eqnarray}
We do not label the elements of the eigenvectors with $M$ because they do not depend on $M$ as shown below.
 Eq.(\ref{Hu})  for $M<N_f$ can be written as  follows:
\begin{eqnarray}\label{u}
\sum_{j=1}^M d_j (u_{m(k-j)} + u_{m(k+j)}) = \lambda^{(M)}_m u_{mk}, 
\end{eqnarray}
which holds for both odd and even $N$.
If $M=N_f$, then we have
\begin{eqnarray}\label{u_odd}
\sum_{j=1}^{(N-1)/2} d_j (u_{m(k-j)} + u_{m(k+j)}) = \lambda^{((N-1)/2)}_m u_{mk},
\end{eqnarray} for odd $N$
and
\begin{eqnarray}\label{u_even}
\sum_{j=1}^{N/2-1} d_j (u_{m(k-j)} + u_{m(k+j)})  + d_{\frac{N}{2}}u_{m(k+N/2)} = \lambda^{(N/2)}_m u_{mk}
\end{eqnarray}
 for even $N$.
Notice that Eq.(\ref{u_odd}) coincides with Eq.(\ref{u}) at $M=(N-1)/2$, while Eq.(\ref{u_even}) can not be considered as a particular case of Eq.(\ref{u}).

We look for the solution to Eqs.(\ref{u}) - (\ref{u_even})  in the form
\begin{eqnarray}\label{uC}
u_{mk}= C_{m1} e^{i k p_m} + C_{m2} e^{-i k p_m}.
\end{eqnarray}
First of all, solution (\ref{uC})  must satisfy cyclic condition (\ref{cycle}) which produces the eigenvectors $u_m$,
\begin{eqnarray}
u_m = (u_{m1}\;\;u_{m2} \;\dots\; u_{mN})^T.
\end{eqnarray}
Then, Eqs.(\ref{u})-(\ref{u_even}) yield eigenvalues $\lambda^{(M)}_m$. Thus, we study the eigenvectors first.

\subsection{Eigenvectors of Hamiltonian}

Substituting (\ref{uC}) into (\ref{cycle}) we obtain
\begin{eqnarray}\label{um0}
C_{m1} \Big( e^{i p_m (N+k)} - e^{i p_m k} \Big) +
C_{m2} \Big( e^{-i p_m (N+k)} - e^{-i p_m k} \Big)=0.
\end{eqnarray}
This equation can be written  as
\begin{eqnarray}
2 \sin\frac{ p_m N}{2} \Big( C_{m1} e^{\frac{ip_m N}{2} + i p_m k} -C_{m2} e^{-\frac{ip_m N}{2}- i p_m k}\Big)=0,
\end{eqnarray}
which must hold for any $k$. Then 
we have
\begin{eqnarray}\label{pm}
\sin\frac{ p_m N}{2} =0 \;\;\Rightarrow \;\; p_m =\frac{ 2 \pi (m-1)}{N},\;\;
m=1,2\dots, N 
\end{eqnarray}
(we consider only nonnegative $p_m$ without loss of generality).
Each $p_m$ generates 2 real linearly independent eigenvectors corresponding to $C_{m2}=\pm C_{m1}$.  
\begin{eqnarray}
\label{um}
u^{(1)}_{mk } = 2 C^{(1)}_{m1} \cos(p_m k),\;\; u^{(2)}_{mk } = 2 C^{(2)}_{m1} \sin(p_m k),
\end{eqnarray}
where $C^{(j)}_{m1}$, $j=1,2$ provide  the normalization for the eigenvectors $u_m^{(j)}$, $j=1,2$:
\begin{eqnarray}\label{umf}
(u_m^{(j)})^\dagger u_m^{(j)} = 1 \;\; \Leftrightarrow \;\; \sum_{k=1}^N (u_{mk}^{(j)})^2=1.
\end{eqnarray}
Since formula (\ref{uC}) includes only two coefficients $C_{m1}$ and $C_{m2}$ for each $m$, two vectors is  the maximal possible number of linearly independent eigenvectors associated with  the fixed $m$.
Although $m$ varies in the range $1\le m \le N$ in the above formulae (\ref{um0})-(\ref{umf}),   
one can show that the set of linearly independent eigenvectors corresponds to $m$ in the restricted range
\begin{eqnarray}\label{meven}
&&
m=1,\dots, \frac{N}{2}+1 \;\;\;{\mbox{for even}} \;\; N,\\\label{modd}
&&m=1,\dots, \frac{N+1}{2} \;\;\;{\mbox{for odd}} \;\; N,
\end{eqnarray}
Let us find  $C_{m1}^{(j)}$ for the cases of even and odd $N$.

\paragraph{$N$ is even.}

If $m=1$ or $m= N/2+1$, then $u^{(2)}_{mk}=0$. Thus we have only one nonzero  eigenvector $u^{(1)}_1$ associated with $m=1$ and one nonzero eigenvector  $u^{(1)}_{(N/2+1)}$ associated with $m=N/2+1$. All other values of $m$ admit two eigenvectors $u^{(1)}_{m}$ and $u^{(2)}_{m}$.
Thus we have $2 (N/2-1) +2 = N$ nonzero linearly independent eigenvectors of form (\ref{um}) with $p_m$ from (\ref{pm}):
\begin{eqnarray}\label{umev1}
u^{(1)}_{mk } &=& 2 C^{(1)}_{m1} \cos\frac{2 \pi (m-1) k}{N},\;\;m=1,\dots, \frac{N}{2}+1,\\\nonumber
u^{(2)}_{mk } &=& 2 C^{(2)}_{m1} \sin\frac{2 \pi (m-1) k}{N},\;\;m=2,\dots, \frac{N}{2},
\end{eqnarray}
in particular,
\begin{eqnarray}\label{umev2}
u^{(1)}_{1k} = 2 C^{(1)}_{11},\;\;\; u^{(1)}_{(N/2+1)k} = 2 C^{(1)}_{11} (-1)^k.
\end{eqnarray}
It follows from normalization (\ref{umf}) and  Eqs.(\ref{umev1}), (\ref{umev2}):
\begin{eqnarray}
C^{(1)}_{m1}&=&C^{(2)}_{m1}= \frac{1}{\sqrt{2N}},\;\;m=2,\dots, \frac{N}{2},\\\nonumber
C^{(1)}_{m1}&=&\frac{1}{2\sqrt{N}},\;\;m=1,\frac{N}{2}+1.
\end{eqnarray}

\paragraph{$N$ is odd.}

If $m=1$, then $u^{(2)}_{mk}=0$. Thus we have only one nonzero eigenvector $u^{(1)}_{1}$ associated with $m=1$. All other values of $m$ admit two eigenvectors $u^{(1)}_{m}$ and $u^{(2)}_{m}$.
Thus we have $2 ((N-1)/2) +1 = N$ non-zero linearly independent  eigenvectors of form (\ref{um}) with $p_m$ from (\ref{pm}):
\begin{eqnarray}\label{umodd}
u^{(1)}_{mk } &=& 2 C^{(1)}_{m1} \cos\frac{2 \pi (m-1) k}{N},\;\;m=1,\dots, \frac{N+1}{2},\\\nonumber
u^{(2)}_{mk } &=& 2 C^{(2)}_{m1} \sin\frac{2 \pi (m-1) k}{N},\;\;m=2,\dots, \frac{N+1}{2}.
\end{eqnarray}
In particular,
\begin{eqnarray}
u^{(1)}_{1k} = 2 C^{(1)}_{11}.
\end{eqnarray}
It follows from normalization (\ref{umf}) and  Eq.(\ref{umodd}):
\begin{eqnarray}
C^{(1)}_{m1}&=&C^{(2)}_{m1}=\frac{1}{\sqrt{2N}},\;\;m=2,\dots, \frac{N+1}{2},\\\nonumber
C^{(1)}_{11}&=&\frac{1}{2\sqrt{N}}.
\end{eqnarray}

We emphasize that all equations derived in this section do not depend on $M$. Thus, eigenvectors of the Hamiltonian do not depend on the number of interacting neighbors $M$.

\subsection{Eigenvalues of Hamiltonian}
Constructing the eigenvalues, we have to consider separately two cases: $M<N_f$ and $M=N_f$.

\subsubsection{$M<N_f$}
Substituting $u_{mk}$ from (\ref{uC}) into (\ref{u}) we obtain
\begin{eqnarray}\label{EV1}
&&
\Big(2\sum_{j=1}^M  d_j \cos(p_m j)  - \lambda^M_m \Big) C_{m1} e^{i p_m k} +\\\nonumber
&&
\Big(2\sum_{j=1}^M  d_j \cos(p_m j)  - \lambda^M_m \Big) C_{m2} e^{-i p_m k} = 0.
\end{eqnarray}
Since this equation must be satisfied for all $k$ we obtain expression for 
$\lambda_m$:
\begin{eqnarray}\label{EV2}
\lambda^M_m = 2\sum_{j=1}^M  d_j \cos(p_m j) , \;\; 1\le M < N_f.
\end{eqnarray}
Set of different eigenvalues corresponds  to $m$ from range (\ref{meven}) for even $N$ and  from range (\ref{modd}) for odd $N$.

\subsubsection{$M=N_f$}

Since (\ref{u_odd}) coincides with (\ref{u}) at $M=(N-1)/2$ the appropriate formulae can be obtained from Eqs.(\ref{EV1}) and (\ref{EV2})  just setting $M=(N-1)/2$, at that $m$ takes the values from the range (\ref{modd}). Therefore we turn to the case of even $N$.

Substituting  $u_{mk}$ from (\ref{uC}) into (\ref{u_even})  and
taking into account that $p_{m} = \frac{2\pi (m-1)}{N}$, i.e.,
\begin{eqnarray}
e^{ ip_m {N/2} }=e^{ -ip_m {N/2} } = (-1)^{m-1},
\end{eqnarray}
we obtain
\begin{eqnarray}\label{EV1even}
&&
\Big(2\sum_{j=1}^{N/2-1}  d_j \cos(p_m j)   + (-1)^{m-1} d_{N/2}- \lambda^{(N/2)}_m \Big) C_{m1} e^{i p_m k} +\\\nonumber
&&
\Big(2\sum_{j=1}^{N/2-1}  d_j \cos(p_m j) + (-1)^{m-1}d_{N/2}  - \lambda^{(N/2)}_m \Big) C_{m2} e^{-i p_m k} = 0.
\end{eqnarray}
Since this equation must be satisfied for all $k$, expression in the parentheses must be zero. Therefore we obtain the expression for 
$\lambda_m^{(N/2)}$:
\begin{eqnarray}\label{EV2even}
\lambda^{(N/2)}_m &=& 2\sum_{j=1}^{N/2-1}  d_j \cos(p_m j)+ (-1)^{m-1}d_{N/2},\\\nonumber
&&
m=1,\dots, \frac{N}{2}+1 
.
\end{eqnarray}

\subsection{Evolution of arbitrary one-excitation state}
Now we can describe the evolution of an arbitrary pure one-excitation initial state $|\psi_0\rangle$. At that,  we use the dimensionless time 
$\tau = \frac{d_1 t}{2}$.
The evolution of the wave vector $|\psi(\tau)\rangle$ is following:
\begin{eqnarray}
|\psi(\tau)\rangle= e^{ -i H^{(M)} \frac{2 \tau}{d_1}} |\psi_0\rangle =Ue^{ -i \Lambda^{(M)}\frac{2\tau}{d_1}} U^\dagger |\psi_0\rangle ,\;\;|\psi_0\rangle =\sum_{j=1}^N a_{j} |j\rangle,\;\; \sum_j|a_j|^2=1,
\end{eqnarray}
where $|j\rangle$ is the state with the $j$th excited spin, the matrices of eigenvalues and eigenvectors 
\begin{eqnarray}\label{evL}
\Lambda^{(M)}&=& {\mbox{diag}}(\Lambda^{(M)}_1, \dots,\Lambda^{(M)}_N),\\\label{evU}
U&=&(u_1\;\;u_2\;\cdots \; u_N), \;\;
u_m = (u_{m1},\dots,u_{mN})^T,
\end{eqnarray}
are defined by their elements in terms of
 $\lambda_m^{(M)}$ and  $u^{(i)}_{mj}$. These matrices  depend on the parity of $N$.

For even $N$ we have
\begin{eqnarray}
\Lambda_{1}^{(M)} = \lambda_1^{(M)},&& \Lambda_2^{(M)} = \lambda_{N/2+1}^{(M)},\;\;
\Lambda^{(M)}_{2 m-1} = \Lambda^{(M)}_{2 m} = \lambda^{(M)}_m,\\\nonumber
&&
m=2,\dots, N/2\\\nonumber
u_{1}=u^{(1)}_{1},&& u_{2}=u^{(1)}_{N/2+1},\;\;
u_{2m-1}=u^{(1)}_{m},\;\; u_{2 m}=u^{(2)}_{m},\\\nonumber
&&m=2,\dots,N/2.
\end{eqnarray}

For odd $N$ we have
\begin{eqnarray}
\Lambda_{1}^{(M)} = \lambda_1^{(M)},&&
\Lambda^{(M)}_{2 (m-1)} = \Lambda^{(M)}_{2 m-1} = \lambda^{(M)}_m,\\\nonumber
&&
m=2,\dots, (N+1)/2\\\nonumber
u_{1}=u^{(1)}_{1},&& 
u_{(2(m-1))}=u^{(1)}_{m},\;\; u_{(2 m-1)}=u^{(2)}_{m},
\\\nonumber
&&
m=2,\dots,(N+1)/2.
\end{eqnarray}

As in \cite{FZ_2023},
the dynamics of an arbitrary initial state is expressed in terms of the probability amplitudes which holds for dynamics of either pure or mixed states ($U$ is a real matrix)
\begin{eqnarray}
p^{(M)}_{jk} = \langle j| e^{ -i H^{(M)} \frac{2\tau}{d_1}}|k\rangle = \sum_n u_{jn}  u_{nk} e^{ -i \lambda^{(M)}_n \frac{2\tau}{d_1}},\;\; j,k=1,\dots,N.
\end{eqnarray}
However, since all nodes are equivalent in the circular chain, we consider only the following set of probability amplitudes:
\begin{eqnarray}
p^{(M)}_{1k}(\tau),\;\; k =1,\dots, N_f,
\end{eqnarray}
where we fix one of the spins assigning it the number 1. 

\section{Analysis of dynamics}
\label{Section:analyt}

\begin{figure*}[!]
\includegraphics[scale=0.4]{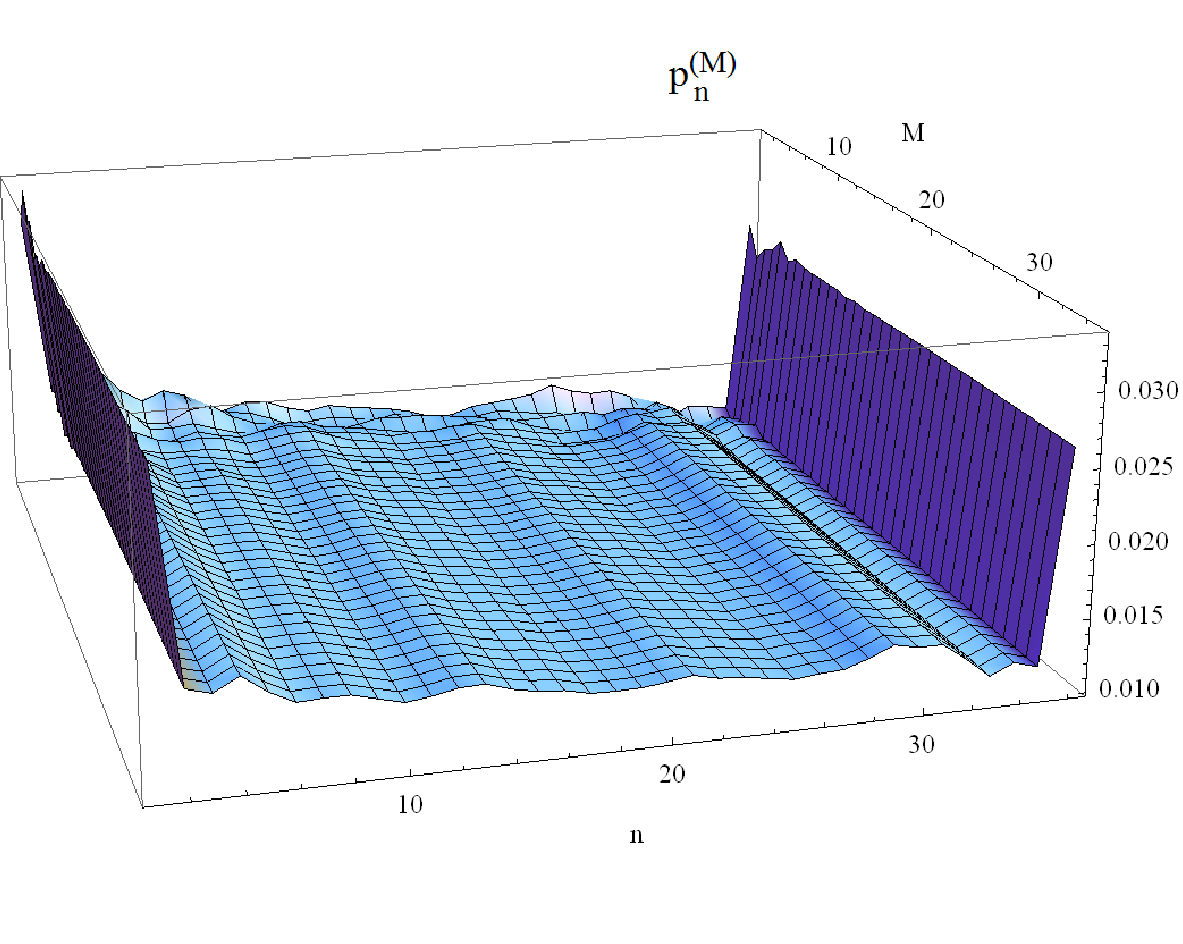}
\caption{The probability $P^{(M)}_n$ averaged 
over the time interval $T=N$ for the chain of $N=70$ spins.}
\label{Fig:P}
\end{figure*}

First, we represent the distribution of the state-transfer probability $|p^{(M)}_{1n}(\tau)|^2$ averaged over the time interval $T=N$, $0\le \tau \le T$,
\begin{eqnarray}
P_n^{(M)} =\frac{1}{T} \int_0^T d\tau  |p^{(M)}_{1n}(\tau)|^2,
\end{eqnarray}
on the plane $(M,n)$ for the chain of $N=70$ spins, Fig.\ref{Fig:P}.
We see on this graph that the behavior of probabilities for small $M\sim 1 \div 10$
significantly differs from the case of large $M$. Remember that all node interaction 
corresponds to $M=N_f=35$ for the $N=70$ node spin chain. The overall behavior of this graph demonstrates decrease of the probability with increase in $n$. The maximum probability correspond to the 1st and 
$(N/2+1)$th spins (which is opposite to the 1st spin). 

\begin{figure*}[!]
\begin{subfigure}[h]{0.45\textwidth}
\includegraphics[scale=0.4]{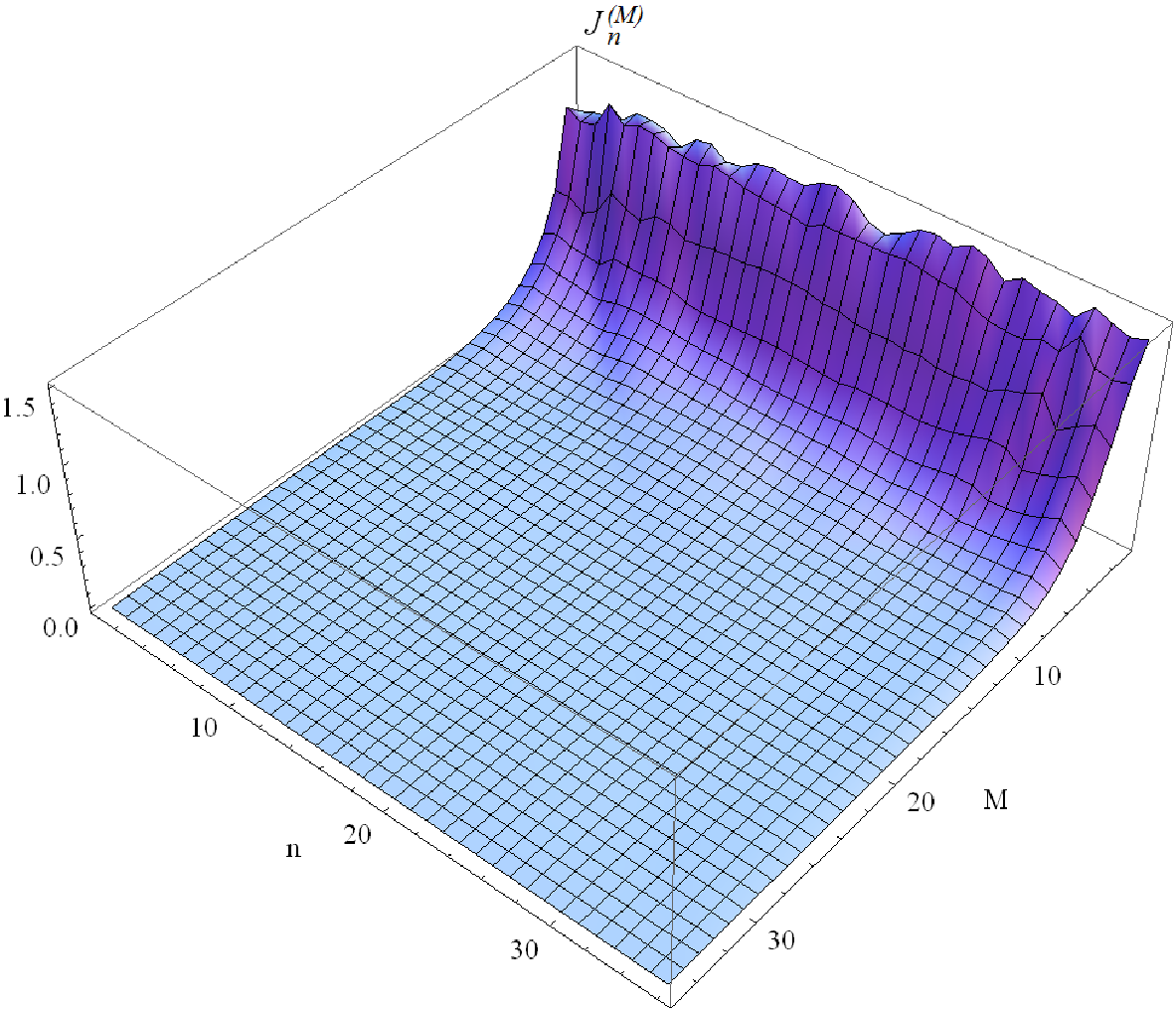}
\caption{}
\end{subfigure}\hfill
\begin{subfigure}[h]{0.45\textwidth}
\includegraphics[scale=0.4]{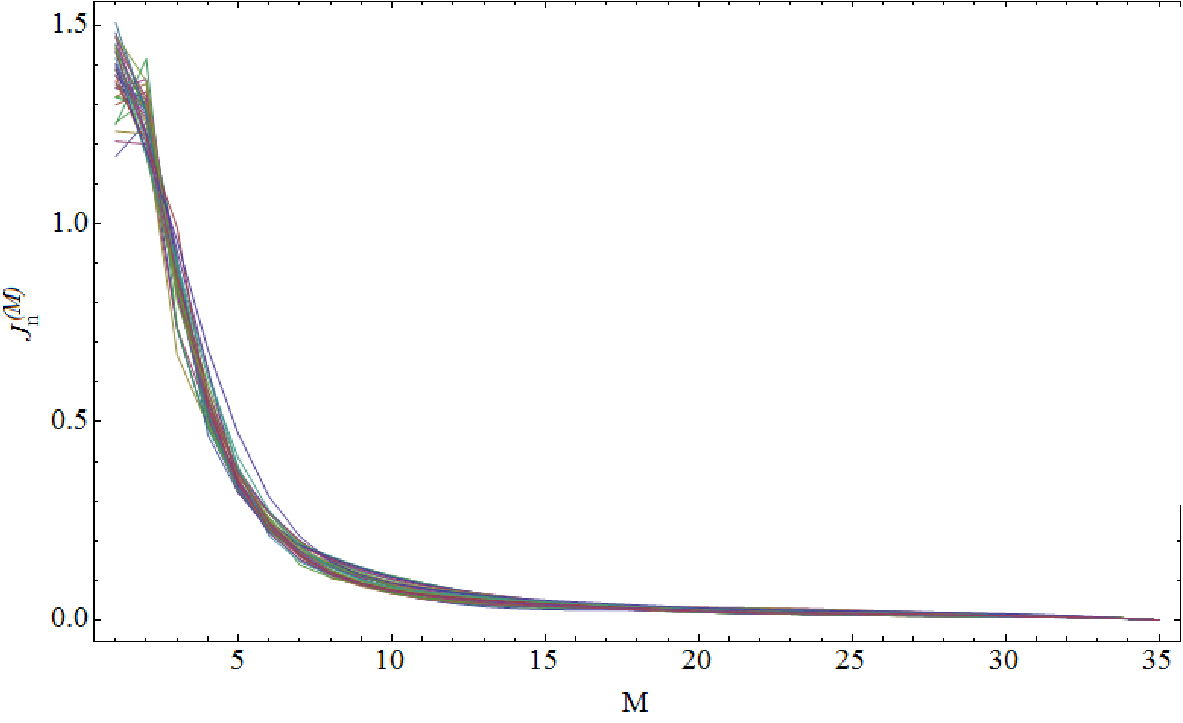}
\caption{}
\end{subfigure}
\caption{The integral characteristics $J^{(M)}(n)$ as functions of 
$M$ and $n$ for the spin chain of $N=70$ nodes. (a) The integral $J^{(M)}(n)$ over the plane $(M,n)$. (b)
The bundle of integrals $J^{(M)}(n)$, $n=1,\dots, 35$, as functions of $M$.
}
\label{Fig:Mn}
\end{figure*}

For the more detailed  analysis of the accuracy of the $M$-neighbor approximation in describing 
the state evolution, we introduce the following integral characteristics \cite{FZ_2023}:
\begin{eqnarray}\label{JMn}
J^{(M)}(n)=\sqrt{\frac{\int_0^T d\tau|p^{(M)}_{n} - p^{(N_f)}_{n}|^2}{\int_0^T d\tau | p^{(N_f)}_{n}|^2}}.
\end{eqnarray}
We take $T=N$.
The parameter $J^{(M)}(n)$  over the plane $(M,n)$ is shown in Fig. \ref{Fig:Mn}a for $N=70$ spin chain.

The bundle of $p^{(M)}_{n}$ for $n=2,\dots, 20$ as function of $M$ is shown in Fig.(\ref{Fig:Mn})b for $N=70$.
It is interesting to note that, unlike the $M$-node approximation for the open chain \cite{FZ_2022},
the integral characteristics $J^{(M)}(n)$ slightly depend on  $n$ due to the symmetry of the system.

Fig. \ref{Fig:Mn}b demonstrates that all curves $J^{(M)}(n)$ as functions of $M$ for different $n$  are very close to each other and the spread of lines reduces with an increase of the time interval $T$  in the definition of $J^{(M)}(n)$ (\ref{JMn}). Fig. \ref{Fig:Mn} shows that small $M \sim 1\div 10$ yield significant error in description of the evolution since the integral is $\gtrsim 0.1$, which agrees with the Fig.\ref{Fig:P} where the shapes of grid lines corresponding to $M\sim 1\div 10$ significantly differ from the shapes of  grid lines  corresponding to $M>10$. As the threshold of accuracy  we  take
such ${\cal{M}}$ that, for the fixed chain length $N$, 
\begin{eqnarray}\label{th}
&&
\forall n \;J^{(M)}(n) \le \varepsilon , \;\; M\ge {\cal{M}},\\\nonumber
&&
\forall n \;J^{(M)}(n)> \varepsilon , \;\; M< {\cal{M}}.
\end{eqnarray}
where $\varepsilon$ is a conventional parameter, we set $\varepsilon=0.1$ \cite{FZ_2022} and represent the obtained ${\cal{M}}$ for different $N$ in the Table \ref{Table:M}.
\begin{table}
\begin{tabular}{|c||ccccccccc|}
\hline
$N$&20& 26& 30& 36& 40& 46& 50& 60& 70\cr
${\cal{M}}$&8& 10& 10& 10& 11&10&10&10& 11\cr
\hline
\end{tabular}
\caption{The parameter ${\cal{M}}$ for the chains of different length $N$.}
\label{Table:M}
\end{table}
We see from this table that ${\cal{M}}=10$ for most lengths $N$. It is likely that ${\cal{M}}$ is a very slowly increasing function of $N$. The fact that ${\cal{M}}=11$ in the middle of this table (for $N=40$) can be explained by the leak of calculation accuracy.

\begin{figure*}[!]
\includegraphics[scale=0.6]{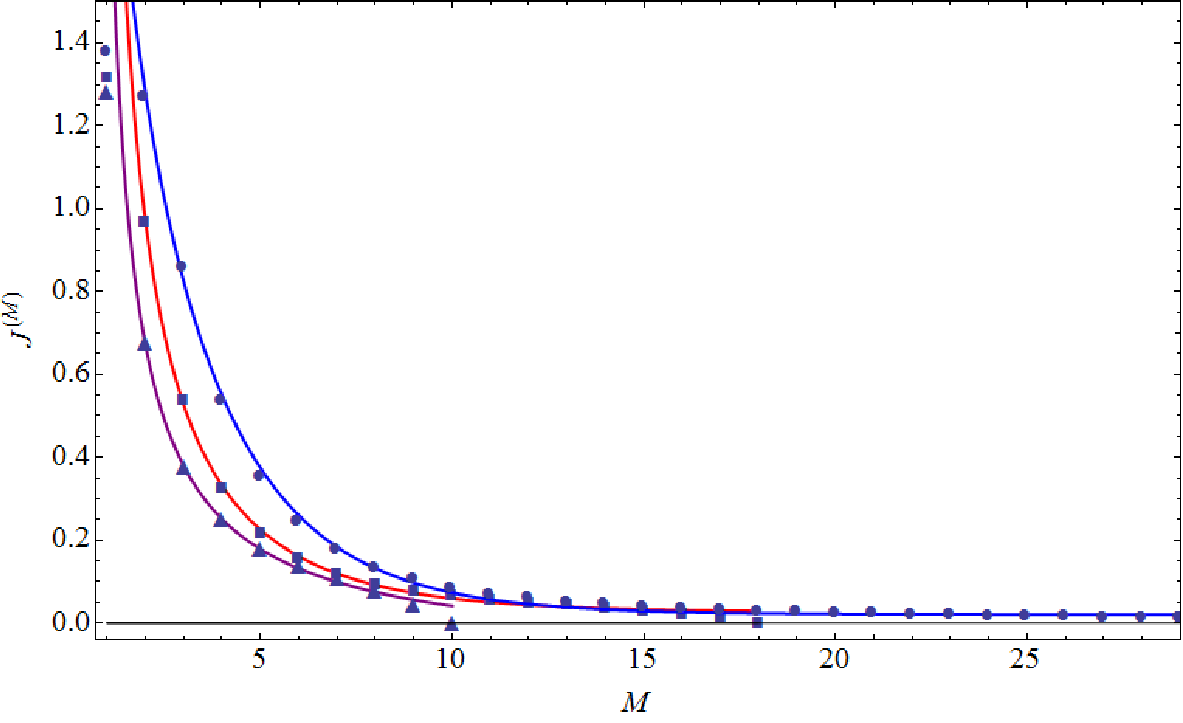}
\caption{The averaged integral $J^{(M)}$ approximated by the curve (\ref{curve}) for $N=20$, $36$ and $70$ (from  bottom to top).}
\label{Fig:avr}
\end{figure*}

For better visualizing, we average $J^{(M)}(n)$ over $n$ obtaining $J^{(M)}$:
\begin{eqnarray}
J^{(M)}&=&\frac{1}{N}\left(2\sum_{n=2}^{(N+1)/2} J^{(M)}(n)+J^{(M)}(1)\right)  \;\;{\mbox{for odd}} \;\; N,\\\nonumber
J^{(M)}&=&\frac{1}{N}\left(2\sum_{n=2}^{N/2} J^{(M)}(n)+J^{(M)}(1) +J^{(M)}(N/2+1)\right)  \;\;{\mbox{for even}} \;\; N,
\end{eqnarray}
where we take into account the parity of the system. 
The result corresponding to chains $N=20$, $36$ and $70$ are shown by  dots on Fig.\ref{Fig:avr}.

\begin{figure*}[!]
\includegraphics[scale=0.6]{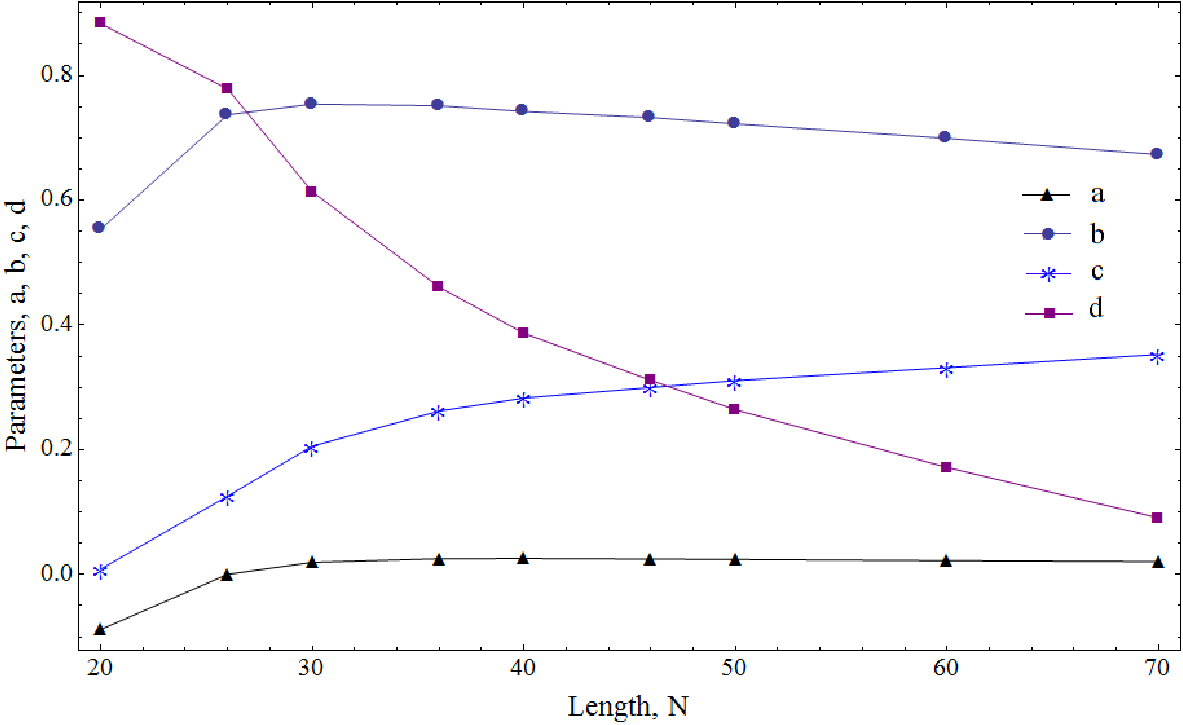}
\caption{Dependence of the parameters $a$, $b$, $c$ and $d$ in Eq.(\ref{curve}) on $N$.}
\label{Fig:abcd}
\end{figure*}

From this figure we can better estimate the number of interacting neighbors $M$ sufficient for describing the evolution  with required accuracy. 
It is interesting, that these points are well approximated over the interval $2\le M \le N_f-1$ by the curve (smooth lines on Fig.\ref{Fig:avr})
\begin{eqnarray}\label{curve}
J=a + \frac{e^{-cx}}{x^d-b}
\end{eqnarray}
We find parameters $a$, $b$, $c$, $d$ for different even lengths $N=20, 26, 30, 36, 40, 46, 50 ,60, 70$, which  is shown by smooth lines in Fig.\ref{Fig:abcd}.

All parameters are slow functions of $N$. The most important parameters are $c$ and $d$, since they characterize the rate of decrease of $J^{(M)}$ with an increase in $N$. The parameter $d$ decreases with $N$. The  parameter $a$ also approaches zero for large $N$.  The parameter $c$  slowly increases while $b$ decreases with $N$. The available data doesn't allow us to state that there are asymptotics  for the parameters $a$, $b$, $c$ and $d$ as $N\to \infty$. 

\section{Conclusions}
\label{Section:conclusions}

We show that the one-excitation block of the $XX$-Hamiltonian for the closed homogeneous spin-1/2 chain can be analytically diagonalized for any allowed number of interacting neighbors. Therefore, the  one-excitation spin dynamics admits analytical description in such model. The feature of the matrix of eigenvectors is that it doesn't depend on the number of interacting neighbors $M$ (which is a consequence of commutativity of the Hamiltonians $H^{(M)}$ with different $M$), unlike the eigenvalues of the Hamiltonian. We also notice that the most eigenvalues are degenerate producing two linearly independent eigenvectors.  

We introduce the integral characteristics showing the accuracy of $M$-neighbor approximation to the spin dynamics with arbitrary one-excitation initial state and reveal such $M$ which provides required accuracy. The above integral characteristics  describe the probability amplitudes for all possible state transfers and thus tests whether  the $M$-neighbor approximation with given $M$ can calculate any probability amplitude with the  required accuracy $\varepsilon$. Due to the symmetry of the circular chain, the number of different state-transfers is reduced  to 
the state transfers  from the single fixed  spin (the 1st spin) to all others.  
Therefore, this characteristics tells us about  accuracy of the $M$-neighbor approximation applied to the dynamics of an arbitrary (either pure or mixed) initial state since such dynamics is expressed in terms of the state-transfer probability amplitudes. 

We show that the integral $J^{(M)}$ as  a function of $M$ can be well approximated by the  curve (\ref{curve}) with parameter $a$ vanishing  for large $N$, the  parameters $c$ and  $d$ (responsible for the rate of growth) are  slowly increasing and decreasing  functions of $N$ respectively. 

Let us emphasize that the integral characteristics $J^{(M)}(n)$ are obtained using averaging over the time interval $T=N$. Therefore, the accuracy of the $M$-neighbor approximation revealed via this integral holds only for $\tau$ restricted by the interval $0\le t \le T$. Beyond this interval, generally speaking, the accuracy will be worse. In other word, the surface on Fig.\ref{Fig:Mn}a and the curves in Fig.\ref{Fig:Mn}b go up with an increase in $T$. 

We represent the analysis of spin dynamics for even $N$. The analysis of the spin dynamics in chains with odd $N$ is quite similar and results are also very similar to those obtained for  even chains. Therefore we do not represent them in this paper. The only remark is that there is no opposite spins in this case (see Fig.\ref{Fig:cycle}), so that  instead of Fig.\ref{Fig:P}, we have a similar figure with only one sharp increase at $n=1$.

We believe that the developed analytical model will be very useful for investigations of 
many-spin dynamics in various problems of magnetic resonance. 

{\bf Acknowledgments} We acknowledge funding from the Ministry of Science  and Higher Education of the Russian Federation (Grant No. 075-15-2020-779).

{\bf Data availability statements.} All data generated or analysed during this study are included in this published article.

{\bf Conflict of interest statement.} The authors declare no conflict of interest.


\begin{thebibliography}{99}

\bibitem{Bose}
 Bose S. Quantum communication through an unmodulated spin chain, Phys. Rev. Lett. 
 {\bf 91}  (2003) 207901
 
 
 \bibitem{CDEL}
 Christandl M., Datta N., Ekert  A., and Landahl A.J. Perfect state transfer in quantum spin networks, Phys.Rev.Lett. {\bf 92} (2004) 187902
 
\bibitem{KS}
Karbach P., and Stolze J. Spin chains as perfect quantum state mirrors,
Phys.Rev.A.  {\bf 72} (2005) 030301(R)
 
\bibitem{GKMT}
 Gualdi G., Kostak V., Marzoli I., and Tombesi P. Perfect state transfer in long-range interacting spin chains,
Phys.Rev. A.   {\bf 78} (2008) 022325



\bibitem{JW}
Jourdan P., Wigner E., Uber das Paulische \"Aquivalenzverbot, Zeitschrift 
f\"ur Physik {\bf 47} (9) (1928)

\bibitem{CG}
H.B.Cruz, L.L.Goncalves, Time-dependent correlations of the one-dimensional isotropic XY
model, J.Phys. C: Solid State Phys. {\bf 14}, 2785 (1981)

\bibitem{Abragam}
 A. Abragam, The Principles of Nuclear Magnetism, Oxford: Clarendon Press. 1961.
 
\bibitem{DPh}
 Ph.J.Davis, Circulant Matrices, Wiley, New York, 1970 
 
 \bibitem{G}
 R.M.Gray,Toeplitz and Circulant
Matrices: A review, Foundations and Trends in Communications and Information Theory: Vol. 2: No. 3, pp 155-239 (2006).
 
 \bibitem{ALC}
V.Engelsberg, I.J.Lowe, J.I.Carolan, Nuclear-Magnetic-Resonance Line Shape of a Linear Chain of Spins, Phys.Rev.B {\bf 7}, 924 (1973)

\bibitem{BFV}
 G.A.Bochkin, E.B.Fel'dman, S.G.Vasil'ev, The exact solution for the free induction decay in a quasi-one-dimensional system in a multi-pulse NMR experiment, Phys.Lett.A {\bf 383},  2993 (2019)

 \bibitem{DMF}
 S.I.Doronin, I.I.Maximov, E.B.Fel'dman, Multiple-quantum dynamics
of one-dimensional nuclear spin systems in solids, J.Exp.Teor.Phys {\bf 91}(3) 597 (2000)
 
\bibitem{CRC2007}
P.Cappellaro, C.Ramanathan, and D.G.Cory,
 Simulations of information transport in spin chains,
Phys.Rev.Lett. {\bf 99}, 250506 (2007)

\bibitem{KKWC}
C.Y.Koh, L.C.Kwek, S.T.Wang and Y.Q.Chong,
Entanglement and discord in spin glass, Laser Physics {\bf 23}(2), 025202 (2013)

\bibitem{YLLF}
B.-L.Ye, B.Li, X.Li-Jost and Sh.-M. Fei, Quantum correlations in critical XXZ system and LMG model, Int.J.Quant.Inf. {\bf 16}, 1850029 (2018)

 
\bibitem{Z_2014}
A.I.Zenchuk, 
Remote  creation of a one-qubit mixed state through a short homogeneous spin-1/2 chain,
Phys.Rev.A {\bf 90}, 052302(13) (2014)
  
\bibitem{ZASO}
Zwick A., \'Alvarez G.A., Stolze J., Osenda O. Robustness of spin-coupling distributions for
perfect quantum state transfer, Phys. Rev. A. {\bf 84} (2011) 022311

  \bibitem{ZASO2}
Zwick  A. ,  \'Alvarez G.A.,
 Stolze J., and Osenda O. 
Spin chains for robust state transfer: Modified boundary couplings versus completely
engineered chains, Phys. Rev. A.  {\bf 85} (2012) 012318

\bibitem{ZASO3}
Zwick A., \'Alvarez G.A. , Stolze J., and Osenda O. Quantum state transfer in disordered
spin chains: How much engineering is reasonable? Quant. Inf. Comput.  {\bf 15}(7-8), (2015)
582
 
 
\bibitem{Yo}
K. Yosida, Theory of Magnetism, Springer, 1996.

 
\bibitem{LL}
Landau L.D., Lifshitz E.M., Quantum mechanics. Non-relativistic theory. Course of theoretical physics, vol.3, Pergamon, New York, 1977.

 
\bibitem{CRC}
P.Cappellaro, C.Ramanathan, and D.G.Cory, Dynamics and control of a quasi-one-
dimensional spin system, Phys.Rev.A {\bf 76}, (2007) 032317





\bibitem{FZ_2022}
 E.B.Fel'dman and A.I.Zenchuk, $M$-neighbor approximation in one-qubit state transfer along zigzag and alternating spin-1/2 chains,
Phys. Scr. {\bf 97}(9) (2022) 095101

\bibitem{FZ_2023}
E.B.Fel'dman and A.I.Zenchuk, Nearest-neighbor approximation in one-excitation state evolution along
  spin-1/2 chain governed by $XX$-Hamiltonian, Physics Letters A {\bf 457} (2023) 128572

  
\bibitem{DFGM}
  S. I. Doronin, E. B. Fel'dman, I. Ya. Guinzbourg, and I. I. Maximov, Supercomputer analysis of one-dimensional multiple-quantum dynamics of nuclear spins in solids, Chem.
Phys. Lett. {\bf 341}, 144 (2001).

\bibitem{DRKH}
V. V. Dobrovitski, H. A. De Raedt, M. I. Katsnelson, and B. N. Harmon, Quantum oscillations without quantum coherence,
Phys. Rev. Lett. {\bf 90}, 210401 (2003).

\bibitem{ZCAPCDRV}
W. X. Zhang, P. Cappellaro, N. Amtler, B. Pepper, D. G. Cory, V. V.
Dobrovitski, C. Ramanathan, and L. Viola, NMR multiple quantum coherences in quasi-one-dimensional spin systems:
Comparison with ideal spin-chain dynamics, Phys. Rev. A {\bf 80}, 052323
(2009).
  
\bibitem{ADLP}
G.A. \'Alvarez, E. P. Danieli, P. R. Levstein, and H. M. Pastawski,
Quantum parallelism as a tool for ensemble spin dynamics calculations, Phys.
Rev. Lett. {\bf 101}, 120503 (2008).
  
\end{thebibliography}
\end{document}